# Fundamental limits in heat assisted magnetic recording and methods to overcome it with exchange spring structures

D. Suess[1], C. Vogler[1], C. Abert[1], F. Bruckner[1], R. Windl[1] L. Breth[1]

[1]Doppler Laboratory, Institute of Solid State Physics, Vienna University of Technology, 1040 Vienna.

*Abstact:* The switching probability of magnetic elements for heat assisted recording is investigated. It is found that FePt elements with a diameter of 5 nm and a height of 10nm show, at a field of 0.5 T, thermally written in errors of 12%, which is significant too large for bit patterned magnetic recording. Thermally written in errors can be decreased if larger head fields are applied. However, larger fields lead to an increase the fundamental thermal jitter. This leads to a dilemma between thermally written in errors and fundamental thermal jitter. This dilemma can be partly relaxed by increasing the thickness of the FePt film up to 30nm. For realistic head fields, it is found that the fundamental thermal jitter is in the same order of magnitude of the fundamental thermal jitter in conventional recording, which is about 0.5 – 0.8 nm. Composite structures consisting of high Curie top layer and FePt as hard magnetic storage layer can reduce the thermally written in errors to be smaller than $10^{-4}$ if the damping constant is increased in the soft layer. Large damping may be realized by doping with rare earth elements. Similar to single FePt grains also in composite structure an increase of switching probability is sacrifices by an increase of thermal jitter. Structures utilizing first order phase transitions breaking the thermal jitter and writeability dilemma are discussed.

1. Introduction

Heat assisted recording is regarded to be the next technology in order to further improve the areal density in magnetic recording. One major issue in heat assisted magnetic recording are thermally written in errors as pointed out by Richter et al. [1]. The origin of this error is the reduced magnetization during writing due to the temperature which is close to the Curie temperature. One possibility to significantly reduce the thermally written in errors is proposed in Ref [2]. It is suggested to use a composition of a soft magnetic material with high Curie temperature exchange coupled to the hard magnetic storage layer such as FePt with a low Curie temperature. A similar structure was proposed by Coffey et al. in order to use the magnetic and thermal gradient in heat assisted recording [3].

In Ref [2] the estimates of improved writeability are done analytically not taking into account the dynamics of the system. The question is if these assumptions and arguments that base on equilibrium thermodynamics are valid in the highly excited states during heat assisted recording.

1. Modeling single FePt

In order to answer this question we present in this letter atomistic simulations using VAMPIRE based on a spin Hamiltonian in order to simulate the heat assisted recording process by integrating the thermal version of the Landau-Lifshitz equation [4]. In particular the switching probability of one magnetic element is calculated as function of the heat pulse peak temperature. The behavior of FePt elements is compared with a composite structure consisting of FePt coupled to a material with a significantly higher Curie temperature such as Fe. In order to allow for fast atomistic simulation of the FePt structure we restrict the exchange interaction to nearest neighbors. The crystal structure is assumed to be simple cubic. We use the exchange interaction strength $J_{ij}$ and the lattice parameter $a$ as fit parameters to reproduce the behavior of the model which takes into account long-ranged exchange interaction [5],[6]. In particular we fitted $J_{ij}$ and $a$ to reproduce magnetization as function of temperature for two different cylindrical particle sizes with diameter and height of 2 nm and 9 nm, respectively [6]. The best fit was obtained for $a$ = 0.24nm *and* $J_{ij}$ = 5.18x10$^{-21}$J. For the atomistic spin moment we use 1.7 $\mu_B$ (corresponds to saturation polarization $J_s$ = 1.43 T) and for the uniaxial anisotropy 9.12 x 10$^{-23}$ J / link (corresponds to uniaxial anisotropy $K_1$ = 6.6 MJ/m³).

Using these parameters the magnetization as function of temperature is calculated for cylindrical elements with a height of 10 nm and various diameters as shown in Fig. 1 for a damping constant of $\alpha$ = 1.0. As investigated in detail in Ref [6] depending on the particle diameter the Curie temperature varies, which causes an additional contribution to the switching distribution in heat assisted magnetic recording.

In order to study the switching probability of one magnetic element for heat assisted magnetic recording we apply a Gaussian heat pulse $T(t)$:

$$T(t) = (T_{max} - T_{min}) \exp\left[-\frac{(t - t_{max})^2}{2\sigma^2}\right] + T_{min} \qquad (1)$$

with $\sigma = 7 \times 10^{-11}$ s, $T_{min} = 270$ K and $t_{max} = 3.1 \times 10^{-10}$ s. The total simulation time is 0.6 ns. An external field of 0.5 T is applied opposing the magnetization.

*Fig. 2* shows the switching probability as function of the peak temperature for a FePt element for various heights of the element. For each temperature the average of 128 simulations is calculated in order to determine the switching probability. In all simulations throughout this paper the damping constant is assumed to be $\alpha = 0.1$ for FePt [7]. It is interesting to note that the final switching probability at high temperature increases with the element height *h,* even up to 30 nm. Hence, in contrast to non-thermal recording, where the thermal stability increases only up to about the domain wall width of the material and a further increase is not beneficial for recording, in heat assisted recording the switching probability can be increased for heigths much larger than the domain wall width. The maximum switching probability is $P_{max} = 0.88$ and $P_{max} = 0.97$ for the film height of 10 nm and 30 nm, respectively. A high switching probability ($P_{max} > 1 - 1\times10^{-4}$) is particularly important for bit patterned media (BPM), where an incorrect reversal directly leads to a bit error. In the investigated samples, even for the largest film height of $h = 30$ nm, 3% of the elements are incorrectly recorded just due to thermal fluctuations. This number is much too high for a reliable BPM. It is worth noting that switching occurs at a temperature (T = 540 K) which is well below the Curie temperature for this particle size as it can be seen by comparing the Curie temperature of *Fig. 1* with the switching probability of *Fig. 2*.

Although $P_{max}$ can be increased for larger elements, even for *d* = 10 nm which allows for patterned elements storage densities of about 4 Tdots/inch², the switching probability is still far away to be close to one. For *d* = 10 nm $P_{max}$= 0.98. Hence, for ultra high density recording media it is important to investigate structures which lead to a higher $P_{max}$ and this for even smaller element diameters.

If the damping constant in the FePt element is increased to $\alpha=1.0$, the switching probability for an element with *d* = 5 nm and *h* = 10 nm can be increased to $P_{max}$ = 94% as shown in *Fig. 3*. However, experimentally it is expected to be a difficult task to increase the damping of FePt without degenerating the anisotropy and broadening the $T_c$ distributions. For low damping ($\alpha$ = 0.02) the expected performance in heat assisted magnetic recording is very low, due to a small maximum switching probability.

Another essential property for magnetic recording is the dependence of the switching probability as function of the element diameter which is shown in *Fig. 4*. It can be seen that for smaller diameter size, $P_{max}$ is reduced. Furthermore, the temperature where 50% of the simulated elements have switched ($T_{50:50}$) is shifted to lower temperature for smaller diameter. This gives an additional contribution to the switching distribution as it will be discussed later in the paper.

The switching probability as function of the external field strength is investigated in *Fig. 5*. As expected the maximum switching probability $P_{max}$ can be increased with higher field strengths. However, a more detailed investigation as discussed in the next paragraph will show that the higher $P_{max}$ sacrifices another important property for magnetic recording.

In heat assisted recording the maximum slope of the switching probability d$P(T)$/d$T$ determines the transition jitter in granular recording and the achievable density in bit patterned magnetic recording. Large values of the maximum slope d$P(T)$/d$T$ are desired as the jitter decreases with increasing d$P(T)$/d$T$. In order to estimate the jitter, we calculate $\sigma_{dP/dT}$, which is obtained by calculating the derivative of P(T). Since the shape d$P(T)$/d$T$ was observed to appear to be similar to a Gaussian function we use this as a fit function. $\sigma_{dP/dT}$ is the standard deviation of the Gaussian function. The smaller $\sigma_{dP/dT}$ the better are the expected jitter values. The fundamental jitter in granular heat assisted recording due to thermal processes during reversal is estimated by

$$\sigma_{x,dP/dT} = \frac{\partial x}{\partial T}\sigma_{dP/dT} \qquad (2)$$

In *Fig. 6* the maximum switching probability $P_{max}$ and $\sigma_{dP/dT}$ are compared. As it can be seen higher head fields increase $\sigma_{dP/dT}$. The origin can be found in the fact that with higher head fields the thermal induced reversal process is shifted to lower temperature. Since the slope of the switching field as function of temperature (d$H_c$/d$T$) is smaller for lower temperature $\sigma_{dP/dT}$ increases at lower temperature. Hence, similar to conventional recording, where the head field strength has to be optimized to write close to the maximum field gradient, also in heat assisted recording exists an optimal write field stremgth. Here, $\sigma_{dP/dT}$ is minimal if d$H_c$/d$T$ is maximal, which is usually the case close to $T_c$. The optimum write field is obtained by the trade-off between maximum $P_{max}$ and minimal $\sigma_{dP/dT}$. The finding of an optimal write field is in contrast to Ref [1], where it is concluded: "…it is shown that storage densities will be limited to 15 to 20 Tbit/in² unless technology can move beyond the currently available write field magnitudes".

In the next section a structure will be presented that leads to high $P_{max}$ values for even smaller head fields, indicating that the a limited head field strength is not the limiting factor in heat assisted recording.

In order to compare the fundamental thermal limits in conventional and head assisted magnetic recording Fig. 7 compares the purely stochastic noise in these both recording schemes. It is important to note that the investigated jitter origins only in the stochastic processes. Perfect non-interacting grains, with no distributions such as $K_1$, easy axis, $J_s$ or $T_c$ are assumed. More precisely the same grain is simulated 128 times under the action of random thermal forces in order to obtain the switching probabilities.

In order to study the thermally induced jitter in conventional recording the switching process of a cylindrical grain with height of 10 nm and a diameter of 8 nm is investigated as function of the field pulse strength. For each data point 128 simulations are averaged. The field pulse is applied at an angle of 10° with respect to the easy axis. The rise time is 0.1 ns and the duration 1ns. A linear rise of the field is assumed until it is kept constant. The maximum field strength was varied from 0 to about 1.2 T. The anisotropy constant is $K_1$ = 0.3 MJ/m³ and the magnetic saturation polarization is $J_s$ = 0.5 T. The damping constant $\alpha$ = 0.02. Due to the finite temperature in the simulations the transition in fields between switching and non-switching is not perfectly sharp. The standard deviation of the switching field linearly increases from $\sigma_{dP/dH}$ = 0.012 T at 100 K to $\sigma_{dP/dH}$ = 0.04 T at 300K. From the standard deviation of the switching field we obtain the jitter as:

$$\sigma_{x,dP/dH} = \frac{\partial x}{\partial H_w} \sigma_{dP/dH} \qquad (3)$$

For a typical head field gradient d$H_w$/d$x$ = 0.05 T we obtain for $\sigma_{x,dP/dH}$ = 0.8 nm. Interestingly, this value of jitter is in a similar order of magnitude as the stochastic jitter in heat assisted magnetic recording ($\sigma_{x,dP/dT}$ = 0.5 nm) although a very different recording scheme is used in these two technologies. These jitter values give a lower limit of obtainable jitter values for perfect grains without any distributions for a given head field gradient and a given thermal gradient in conventional recording and heat assisted recording, respectively. It is worth noting that in the conventional recording the maximum switching probability for sufficient high fields is basically one. For the used head fields in the heat assisted process, the maximum switching probability is $P_{max}$ = 0.88. Hence, concerning the fundamental limit there is not a significant advantage due to heat assisted recording, which could possibly decrease the jitter by an order of magnitude, as it might be concluded from the huge effective head field gradient in heat assisted recording, which is larger by order of magnitude as in conventional recording.

## 2. Modeling FePt coupled to high $T_c$ material and comparison with single FePt

In order to improve the maximum switching probability we investigate in this section the influence of a high $T_c$ material on the recording process. In particular we investigate the switching performance of a bilayer structure where 5 nm are FePt and 5 nm are a Fe like alloy with a Curie temperature which is 40% higher than that of FePt. Fig. 8 shows the switching probability of a bilayer structure with a magnetic moment which corresponds to 1.4 T. The red curve shows the results for a bilayer structure, where the high Curie temperature layer (Fe) has the same damping constant as FePt ($\alpha$ = 0.1). Due to the slow relaxation to the reversed state in the case of low damping the switching probability of the bilayer structure is worse than that of the single FePt element. However, a significant improvement of the switching probability is obtained as the damping constant in Fe is increased to $\alpha$ = 1.0. Whereas it seems to be very challenging to increase the damping constant in FePt without degenerating the anisotropy constant, a significant increase of the damping constant in soft magnetic elements, such as NiFe can be realized by rare-earth dopants [8].

In Fig. 8 the switching probability of the single FePt element is compared with two bilayer structures with different saturation magnetization in the high $T_c$ material. In both structures the damping constant in the high $T_c$ layer is $\alpha$=1.0. If the saturation magnetization is increased from $J_s$ = 1.4 T to $J_s$ =2.1 T the switching probability further increases. Above $P_{max}$=537 all performed simulations, which are 1152, lead to successful switching.

The reversal process for the bilayer with $J_s$ = 1.4 T is shown in *Fig. 9*. The z-component of the magnetic spins on each atom is color coded. Initially the particle is magnetized up. The arrows at the right of each image show the average magnetic moment of the corresponding layer. At a temperature of 700K the average magnetization in the bottom layer vanishes since the temperature is well above the Curie temperature. Due to the fact that the Curie temperature of the top layer is about 40% higher, even at this elevated temperature the average magnetization in this layer is still 39% of the saturation value. As a consequence of the high magnetization the thermally written in errors are reduced.

The influence of switching probability as function of the diameter for the bilayer media is shown in *Fig. 10*. It can be seen that for the bilayer with a diameter $d$=4 nm not all performed simulations lead to successful switching ($P_{max}$ = 0.98). As $J_s$ is increased to $J_s$ = 2.21 T the switching probability can be increased to $P_{max}$ = 1 - 1.7 x $10^{-3}$. The simulations show that the temperature difference between adjacent islands has to be larger than 120K during writing. Hence, if the near field transducer focuses the energy sufficiently to obtain this temperature gradient, islands of $d$ = 4 nm can be reliably written, which corresponds to an areal density of about 25 Tdots/inch².

The key parameters for various media are compiled in Table. 1. The parameters to characterize various media are: (i) the temperature at which 50% of the elements have switched, which is denoted with $T_{50:50}$ (ii) the switching probability $P_{max}$, which is the probability that the element has switched at a temperature at which the switching probability is the maximum (iii) The maximum slope of the P(T) curve, which is denoted with $(dP(T)/dT)_{max}$ and (iv) the standard deviation of $dP(T)/dT$ which is $\sigma_{dP/dT}$.

From Table. 1 it can be seen that the structure "bilayer 2" shows very good performance. It shows a switching probability larger than $P_{max}$ = 1 - $10^{-4}$ which is significantly better than for the single phase element "FePt 1" which has the same dimensions. The good performance of the structure "bilayer 2" is due to (i) the large magnetization in the top layer, (ii) the high damping in the top layer and (iii) the enhanced Curie temperature of the top layer.

A disadvantage of all investigated bilayer structures is the smaller maximum slope of the $P(T)$ curve $(dP(T)/dT)_{max}$ which directly relates to a larger standard deviation of $\sigma_{dP/dT}$, which increases the fundamental thermal jitter $\sigma_{x,dP/dT}$.

Taking into account the temperature gradient $dT/dx$ = 18 K/nm in the media [9],[10], we can estimate the contribution of position jitter which is introduced due to the fundamental thermal jitter. For the case of the single phase FePt element with a diamter of 5 nm ($\sigma_{dP/dT}$ =9.5 K) one obtains $\sigma_{x,dP/dT}$ = 0.5 nm. For the bilayer element (bilayer 2 of Table 1) with $\sigma_{dP/dT}$ =21.5 K one obtains $\sigma_{x,dP/dT}$ = 1.19 nm.

From *Fig. 4* an *Fig. 10* it is possible to extract the change of $P_{50:50}$ as function of the grain size. The data is summarized in Fig. 11. It is shown that for both, the single FePt element as well as for the bilayer $P_{50:50}$ increases with grain size. More importantly the slope of $dP_{50:50}/dT$ increases with decreasing grain size, which implies that fluctuation variations in the grain size lead to an more pronounced increase of jitter for small grain sizes. A change in diameter of FePt from 4 nm to 5 nm leads to a variation of the $T_{50:50}$ from 510 K to 524 K. A further increase from 5 nm to 10 nm increases $T_{50:50}$ = 541 K. If we assume a grain size (diameter) distribution of 16 % [11], we obtain in the region of $d$=4 to $d$=5 nm, a difference in the write temperature of about 11K. Under the assumption of $dT/dx$ = 18 K/nm this relates to a jitter due to 16% grain size variation of about $\sigma_{x,d}$ = 0.6 nm. It is worth noting that this jitter contribution is in the same order of magnitude as the fundamental thermal jitter. For the bilayer element a change in grain size of 16% leads to a jitter contribution of $\sigma_{x,d}$ = 0.93nm. The origin of a larger $dP_{50:50}/dT$ for the same grain size for the bilayer can be found in the fact that the amount of hard magnetic FePt phase is smaller for the bilayer. As it can be concluded from Fig. 11 $dP_{50:50}/dT$ increases with decreasing volume of FePt phase. Hence, for the same grain the volume of FePt is smaller for the bilayer and hence $dP_{50:50}/dT$ larger.

For the single FePt grain as well as for the bilayer structure a variation of the Curie temperature linearly changes $P_{50:50}$. Hence, a variation of the Curie temperature of 2% leads to the same jitter as the fundamental thermal jitter at a grain size of 5 nm. It is important to note that in measurements of the Curie temperature distribution as reported in Ref [xxxx], the contribution of the fundamental thermal jitter is already included. Hence, even for identical grains without any distributions the transition from switching to non-switching in the experiment would not be perfectly sharp. If the Curie temperature distribution measurements were performed on a time scale as reported here (~0.5 ns) even for a perfect media with all grains having the same intrinsic Curie temperature the expected measured $T_c$ distribution for $d$=5 nm would be around 2%.

The estimates of the jitter values discussed in this section are summarized in Table. 2.

## 3. Conclusion and outlook

To conclude it is shown that for heat assisted recording the head field strength has to be optimized in order to obtain a compromise of sufficient high maximum switching probability $P_{max}$ and a sufficient small jitter. A too low head field leads to too low $P_{max}$. A too high head field leads to too high jitter ($\sigma_{x,dP/dT}$) due to a small slope of the d$P$/d$T$ loop, which originates from switching at a temperature well below the Curie temperature.

If the optimal head field cannot be achieved with current heads, a composite structure where the top layer has (i) a high Curie temperature, (ii) a large magnetic moment and a (iii) high damping constant significantly improves the maximum switching probability. Similar to the single FePt grain this advantage is sacrificed by an increase of the fundamental thermal jitter contribution $\sigma_{x,dP/dT}$ and a stronger dependence of the switching temperature $T_{50:50}$ as function of element diameter.

Hence, the ultimate goal are structures that optimized at the same time $P_{max}$ and $\sigma_{x,dP/dT}$. For the concept of heat assisted recording a high $\sigma_{x,dP/dT}$ is expected if $P_{50:50}$ is close to the Curie temperature of the hard magnetic phase. Close to the Curie temperature the magnetic properties such as $K_1$ and $H_c$ are quickly changes with temperature leading to a large slope of the d$P$/d$T$ loop.

One idea of shifting $P_{50:50}$ for the high $T_c$ / Low $T_c$ structure closer or even exactly to $T_c$ is to increase the anisotropy in the softer high $T_c$ material (e.g. by using CoCrPt alloys as top layer). Due to the high $T_c$ in the top layer the anisotropy constant in the high $T_c$ material is still large if $T_c$ of FePt is applied. Hence, tailoring the magnetic properties allow to design a media, which switched at a temperature of the Curie temperature of FePt but still has a large magnetic saturation magnetization at this temperature. Atomistic simulations however, showed that such a tailored material overall does not show a benefit. The reason is that the basically constant anisotropy constant as function of temperature of the top layer at $T=T_c$ leads to an overall to small change of the effective anisotropy consatant $K_{eff} = (K_{top} + K_{bottom})/2$ as function of temperature. The consequence are again too small slopes d$P$/d$T$.

Another concept which inherently solve the dilemma of having small $P_{max}$ and large $\sigma_{x,dP/dT}$ are exchange spring structures that relies on first order phase transitions. In contrast to the Curie temperature which is a second order phase transition and per definition does not show an infinite fast change of the magnetization as function of temperature, first order phase transitions show a jump at the critical temperature. In Ref **Error! Reference source not found.** the transition temperature of a 50nm thick FeRh film was investigated where the transition from non-magnetic to magnetic occurs at a temperature of $T$ ~ 345 K with a standard deviation of approximately 3-4K. In Ref [14] the transition of Ni-Mn-Sn-Co shape memory alloys occurs with a standard deviation of about 2-3 K. Depending on the Co content the transition temperature can be widely varied. The transition from magnetic to non-magnetic with increasing temperature occurs for $Ni_{52}Mn_{25}Ga_{23}$ with a standard deviation of about 1-2 K at about $T$ = 358 K [15]. Both transitions non-magnetic to magnetic as well magnetic to non-magnetic can be used for heat assisted recording. As described by Thiele et al. the exchange spring effect using FeRh can be utilized for heat assisted recording [16]. As illustrated in Fig. 13 (left) the grains consist of a hard magnetic bottom layer (e.g FePt) and a softer layer which shows a first order phase transition. In the region, where the temperature is below the critical temperature the grains cannot be switched due to the lack of write assist of the top layer, which is in the non-magnetic state. As the temperature increases the top layer becomes ferromagnetic, acting as a write assist layer. The significant advantage is that, the transition from non-magnetic to magnetic in principle can occur at temperature regions as small as 1-2 K leading to extremely small jitter values and at the same time due to the large magnetization in the soft layer also to large $P_{max}$. If the first order transition temperature of the top layer does not depend on the grain size, the structure shows also an insensitivity of the switching temperature on grain size variations.

It is worth noting, that also phase transition from magnetic to non-magnetic can be used for heat assisted recording. Here, the effect is reversed. The top layer of the heated grains becomes non-magnetic. Hence, the heated grains are protected from switching. Whereas the grains below the critical temperature of the top layer are reversed due to the exchange spring effect.

A further possibility to apply first order phase transition is to use an exchange spring structure with a soft and a hard layer, which are coupled (coupling layer) by a phase change material such as FeRh in order to switch on/off the exchange as function of temperature. If the sample is heated the soft and hard layer are exchange coupled leading to a significant reduction of the switching field, which allows to reverse the grain. In the region where the temperature is below the critical temperature of the coupling layer the hard layer cannot be reversed due to the lack of the write assist of the soft layer.

Due to the possibility to tailor the critical temperature of shape – memory alloys and magneto caloric materials over a wide range the requirements for the near field transducer can be significantly relaxed. The critical temperature can be adjusted just above the allowed operational temperature of the hard disk.

The support from the CD-laboratory AMSEN (financed by the Austrian Federal Ministry of Economy, Family and Youth, the National Foundation for Research, Technology and Development), from the FWF – SFB project F4112-N13, and the support from Advanced Storage Technology Consortium (ASTC) is acknowledged. The computational results presented have been achieved using the Vienne Scientific Cluster (VSC).

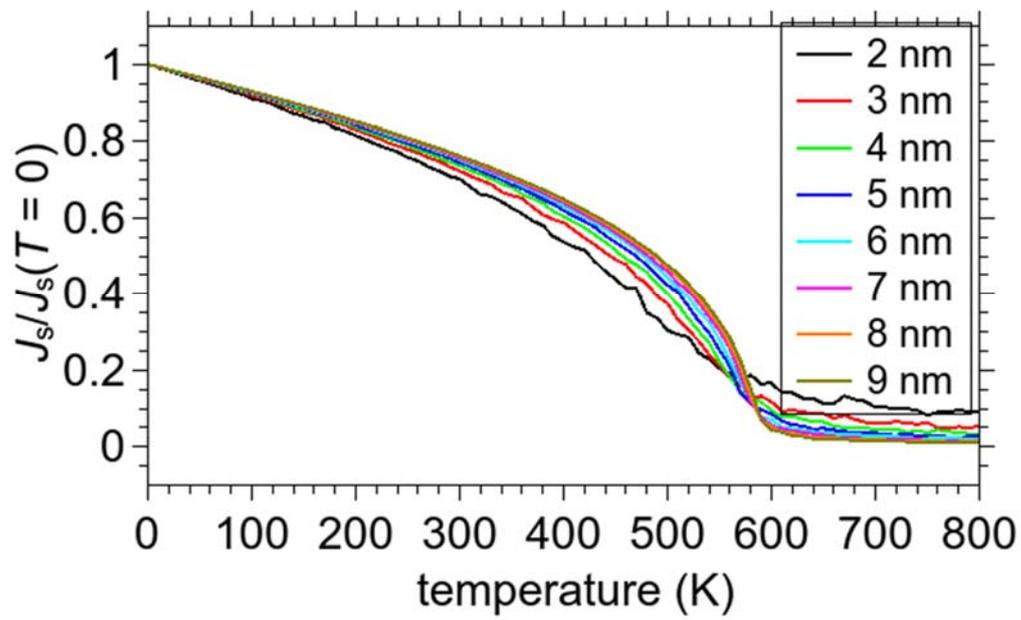

Fig. 1: Magnetization as function of temperature of FePt element for various particle diameter and *h* = 10 nm

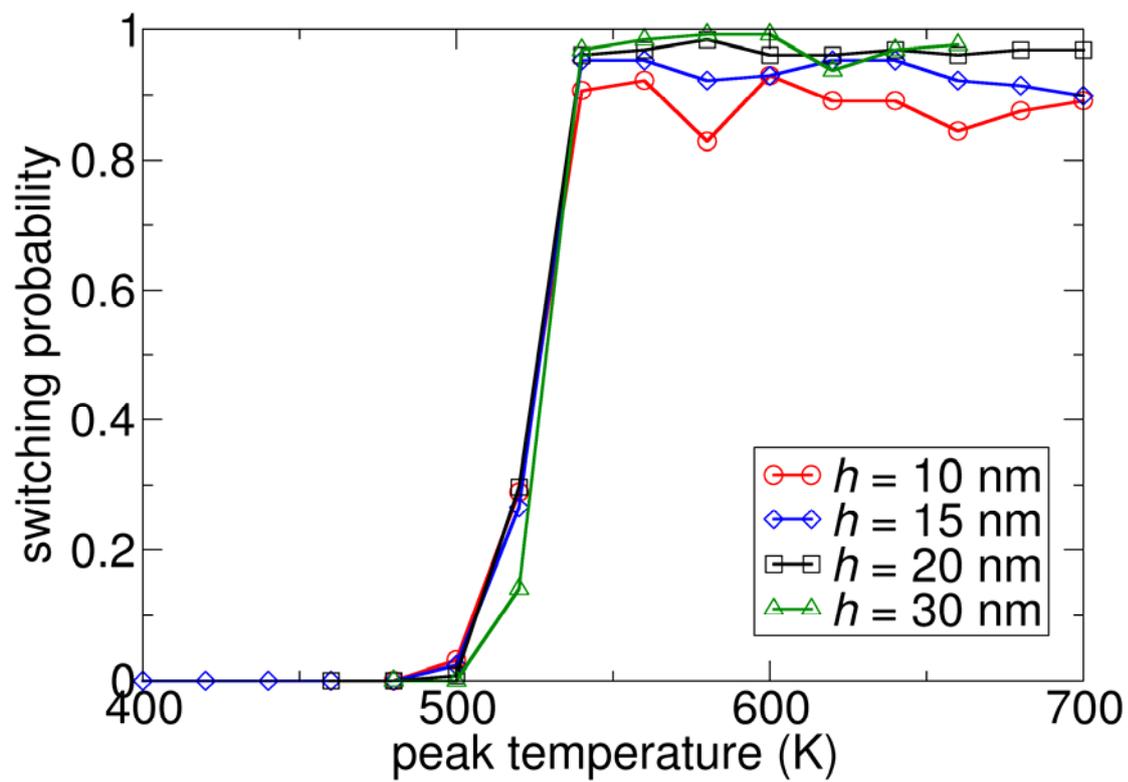

Fig. 2: Switching probability (*P*(*T*) –curve) for a single FePt element for different element heights.

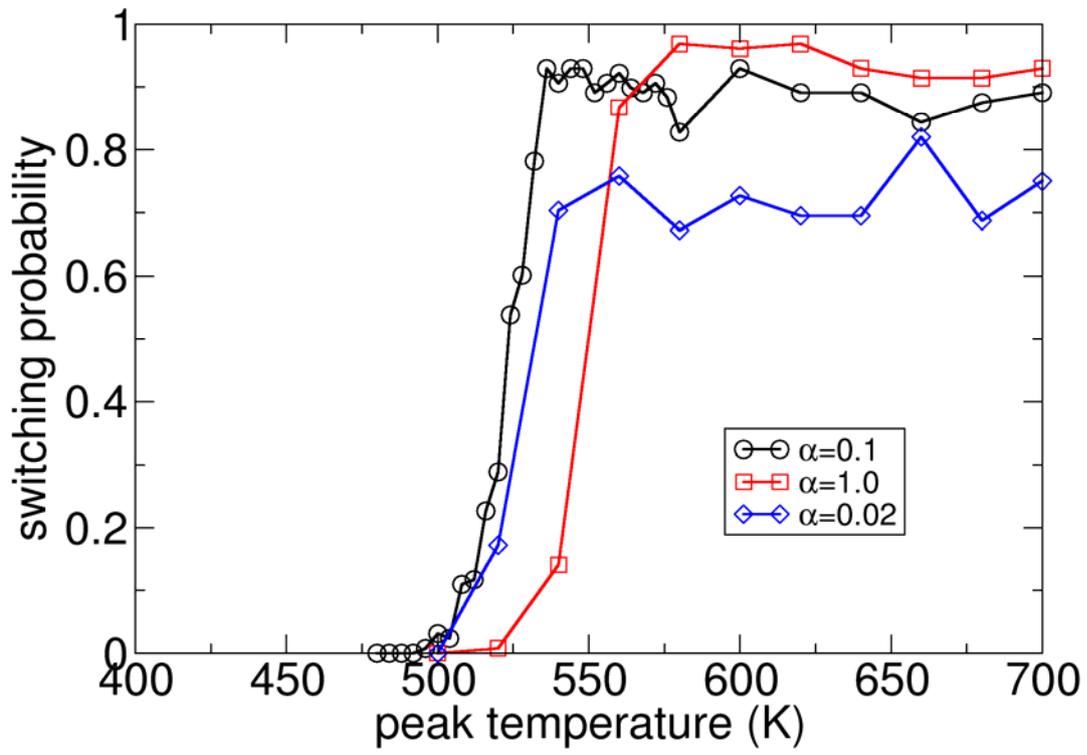

Fig. 3: Switching probability ($P(T)$ –curve) for a single FePt element for different element damping constant in FePt.

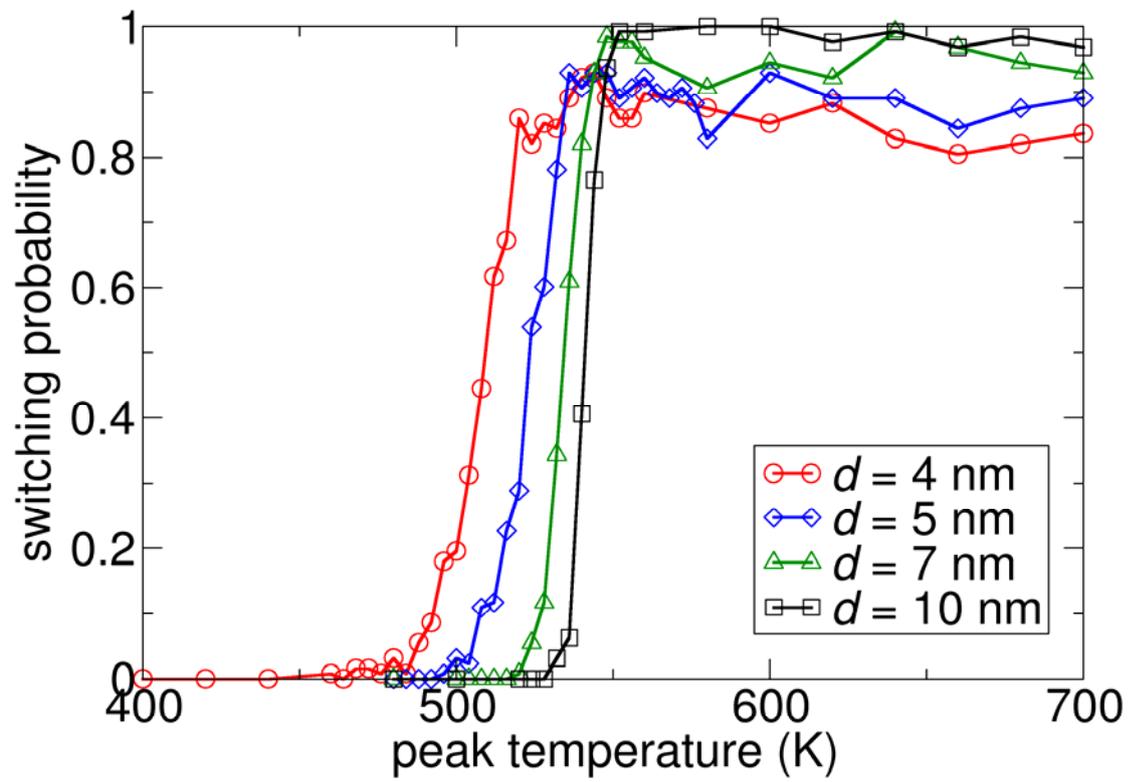

Fig. 4: Switching probability ($P(T)$ –curve) for a single FePt element as function of the diameter *d*.

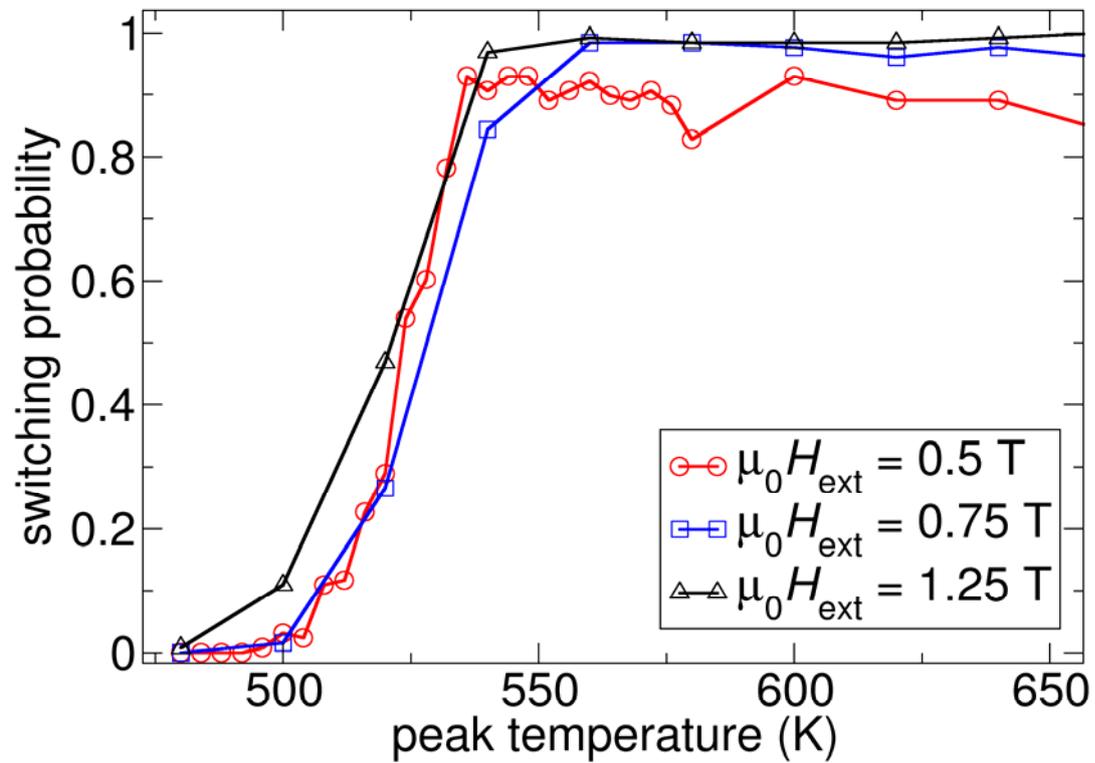

Fig. 5: Switching probability (*P*(*T*) –curve) for a single FePt element as function of the external applied field.

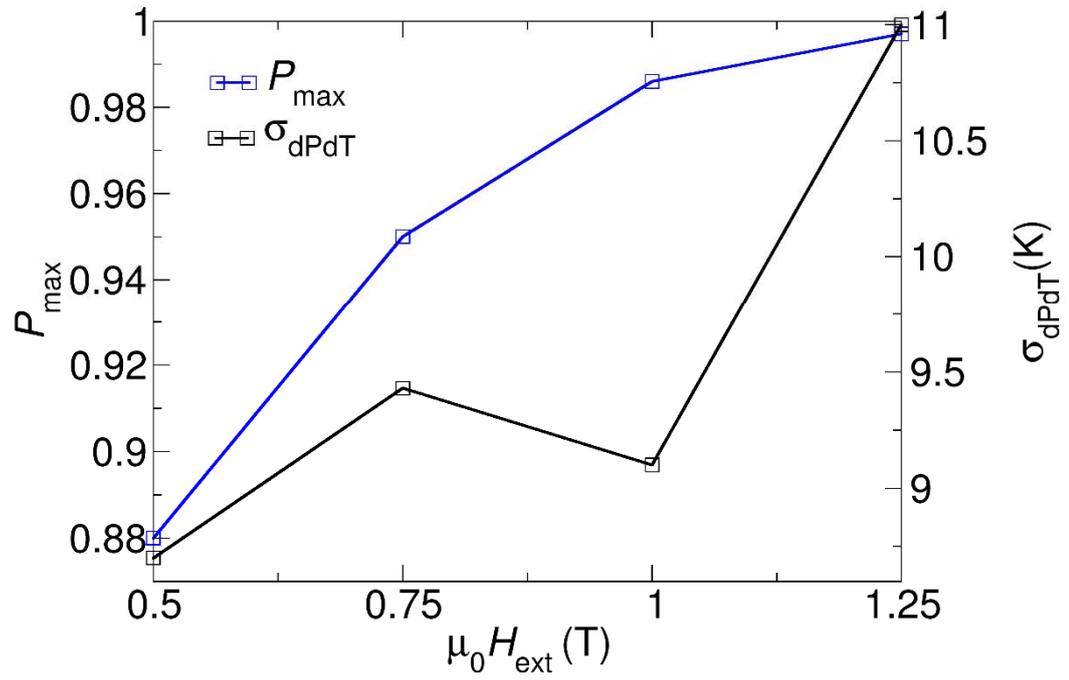

Fig. 6: Comparison of $P_{max}$ and $\sigma_{dPdT}$ for FePt for different strength of the external applied field $\mu_0 H_{ext}$.

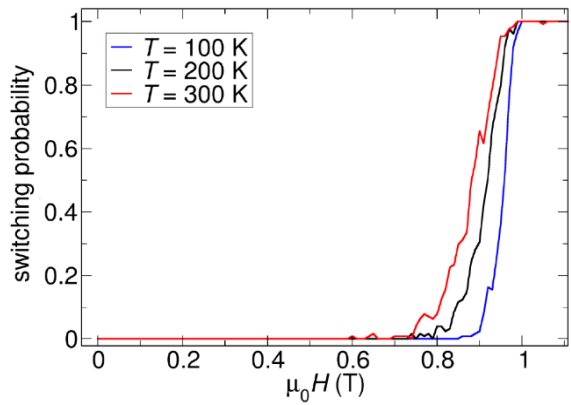 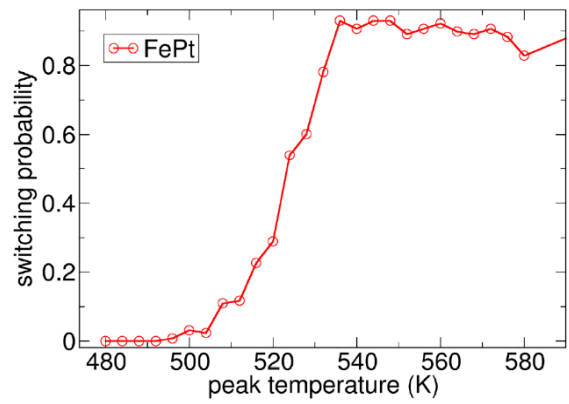

- $\sigma_{dP/dH}$ = 0.04 T  (~ 5% of $H_c$)
- Head field gradient = 0.05 T/nm:
- $\sigma_{x,dP/dH}$ = 0.8 nm

- $\sigma_{dP/dT}$ = 9 K  (~ 2% of $T_{50:50}$)
- Thermal gradient = 20 K/nm:
- $\sigma_{x,dP/dT}$ = 0.5 nm

Fig. 7: Comparison of fundamental thermal jitter (no distributions in anisotropy, grain size and Curie temperature) of conventional perpendicular recording and heat assisted magnetic recording.

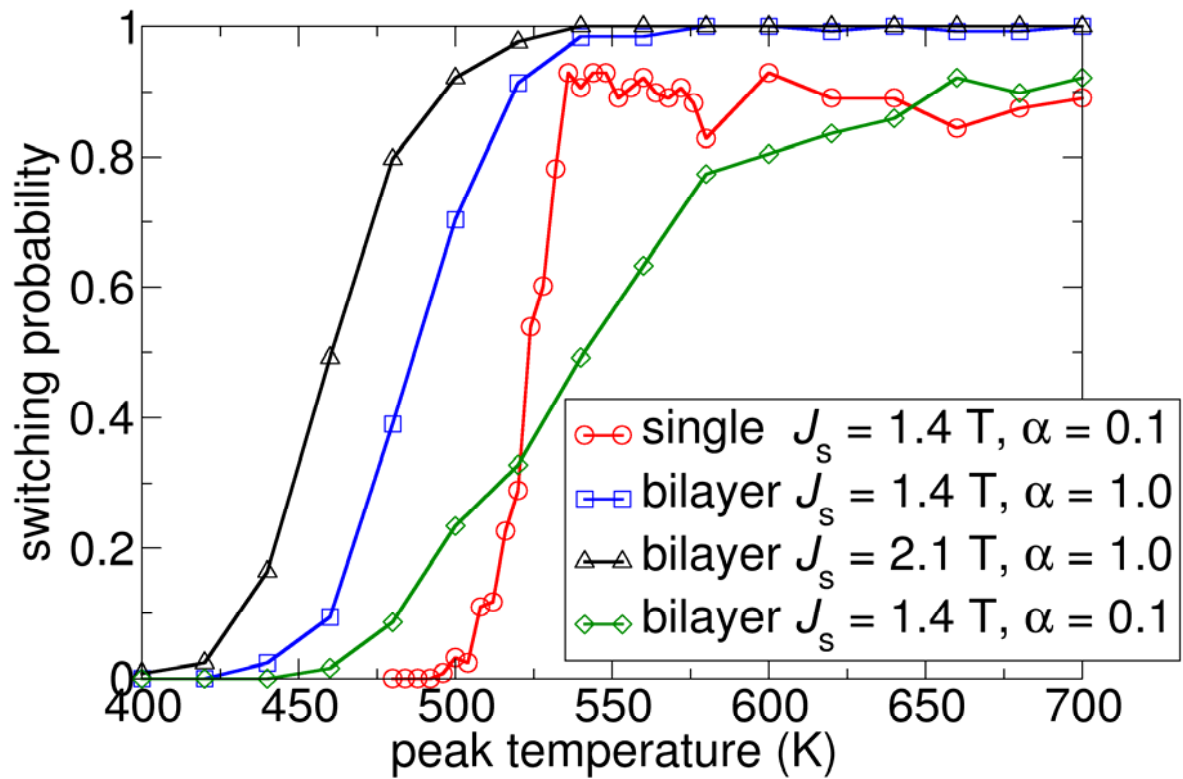

Fig. 8: Switching probability (*P*(*T*) –curve) for bilayer as function of the damping constant and different saturation magnetization in the top layer.
*d* = 5 nm. h = 10 nm

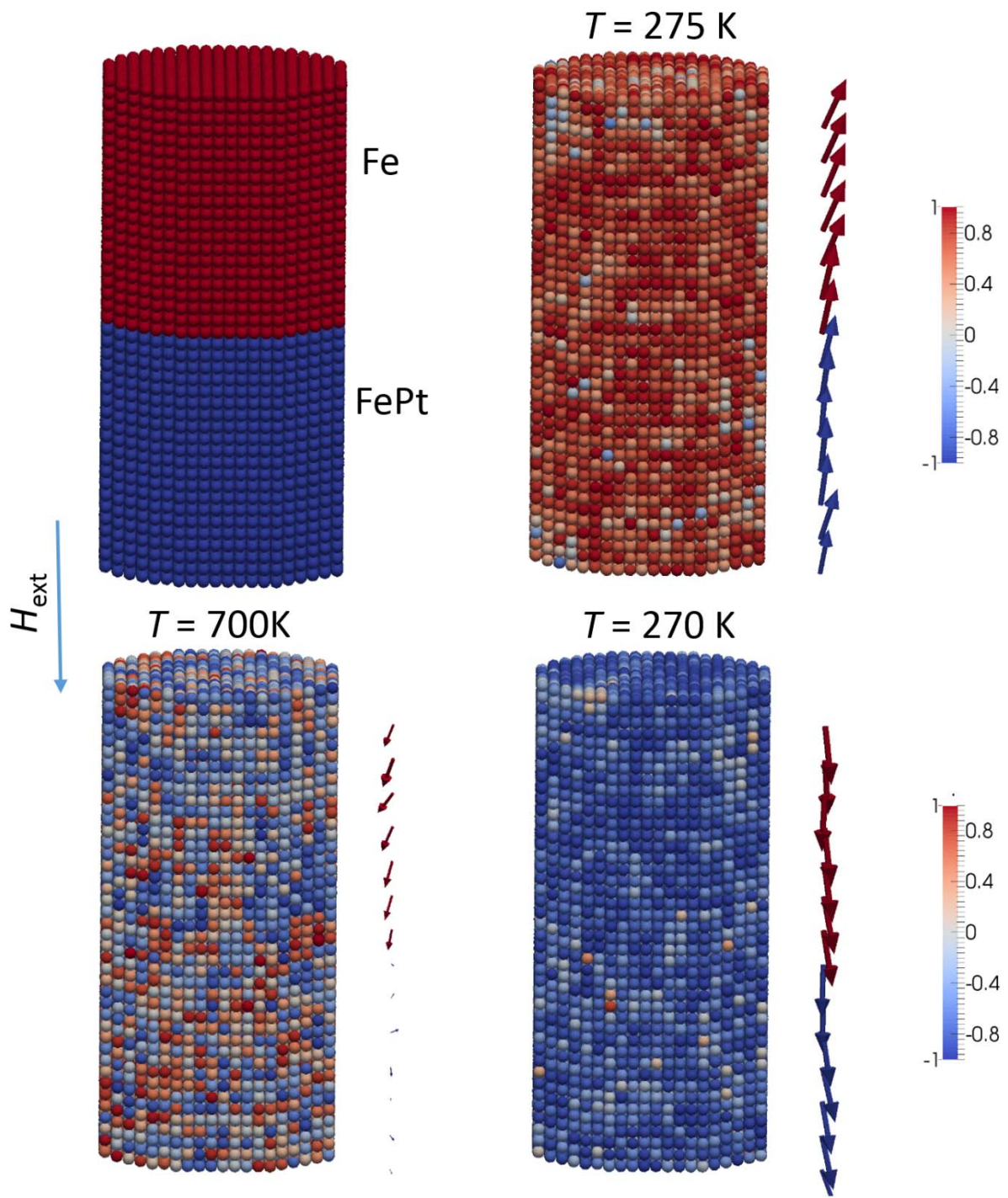

Fig. 9: (left top) atomistic spin position and material distribution of the modeled structure. The z-component of the magnetic spin is color coded. The arrows indicate the average magnetization in the corresponding layer.

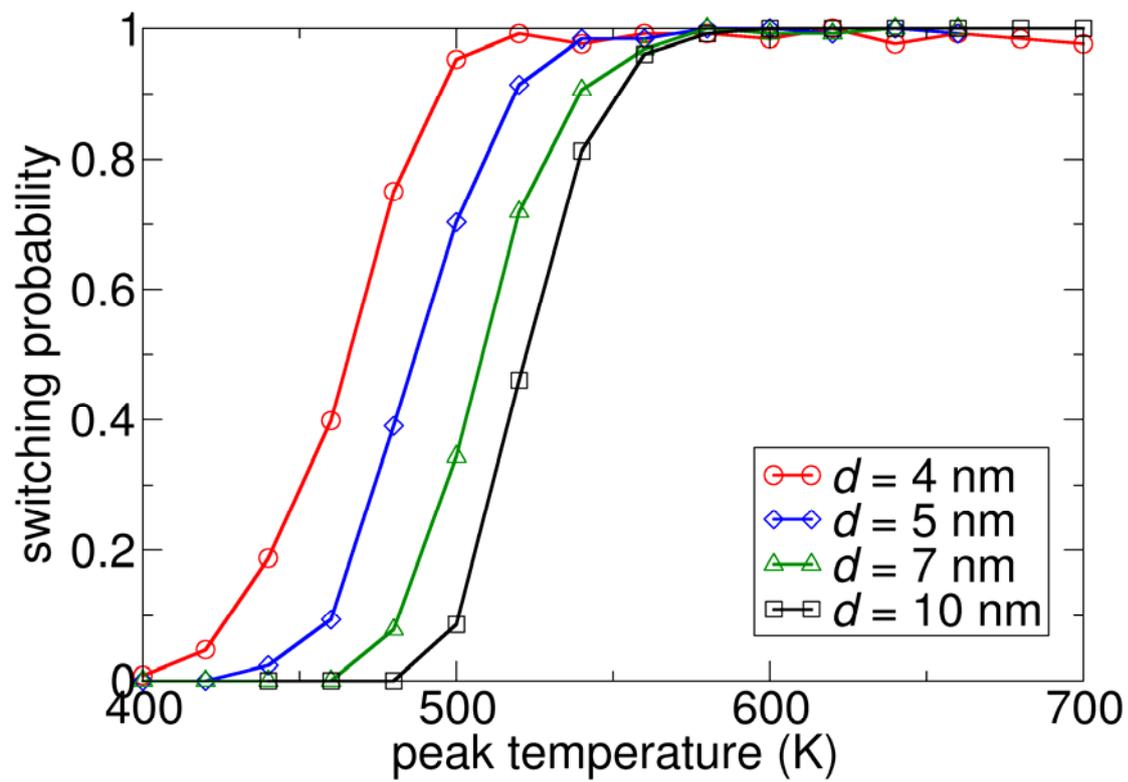

Fig. 10: Switching probability ($P(T)$ –curve) for a bilayer as function of diameter. $\alpha_{top}$ = 1.0.

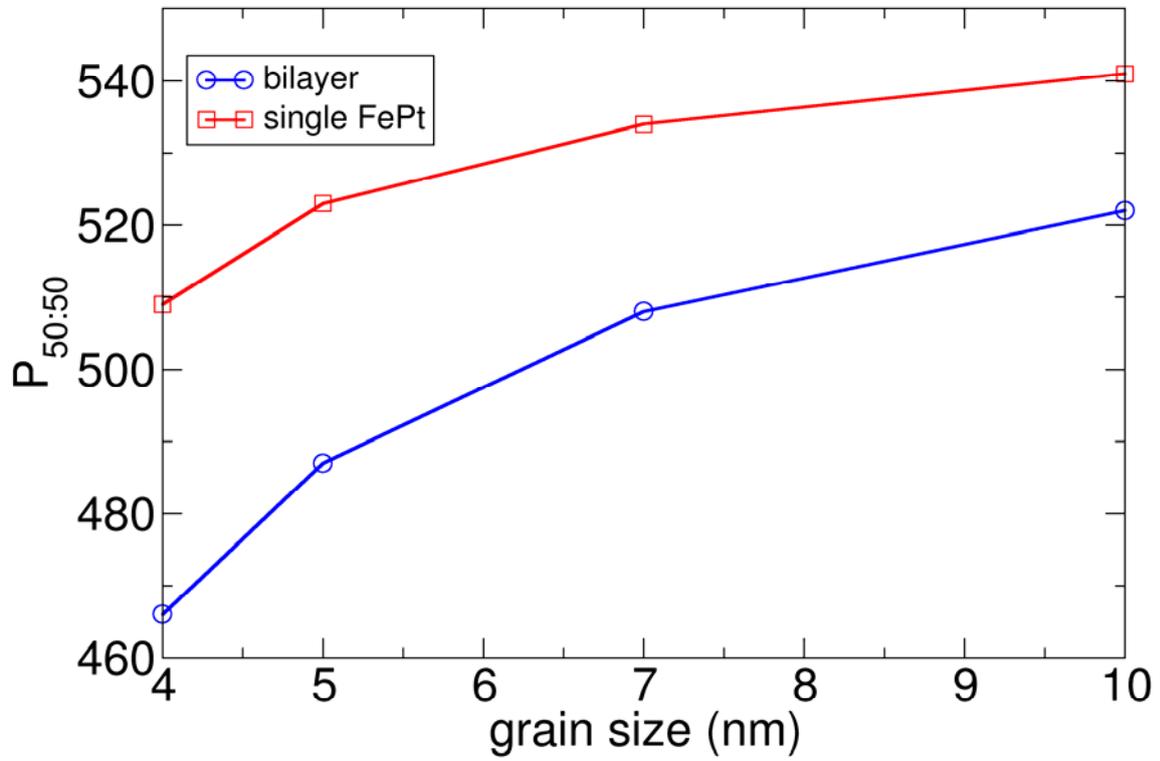

Fig. 11: Comparison of grain size dependence of single FePt grains and bilayer grains.

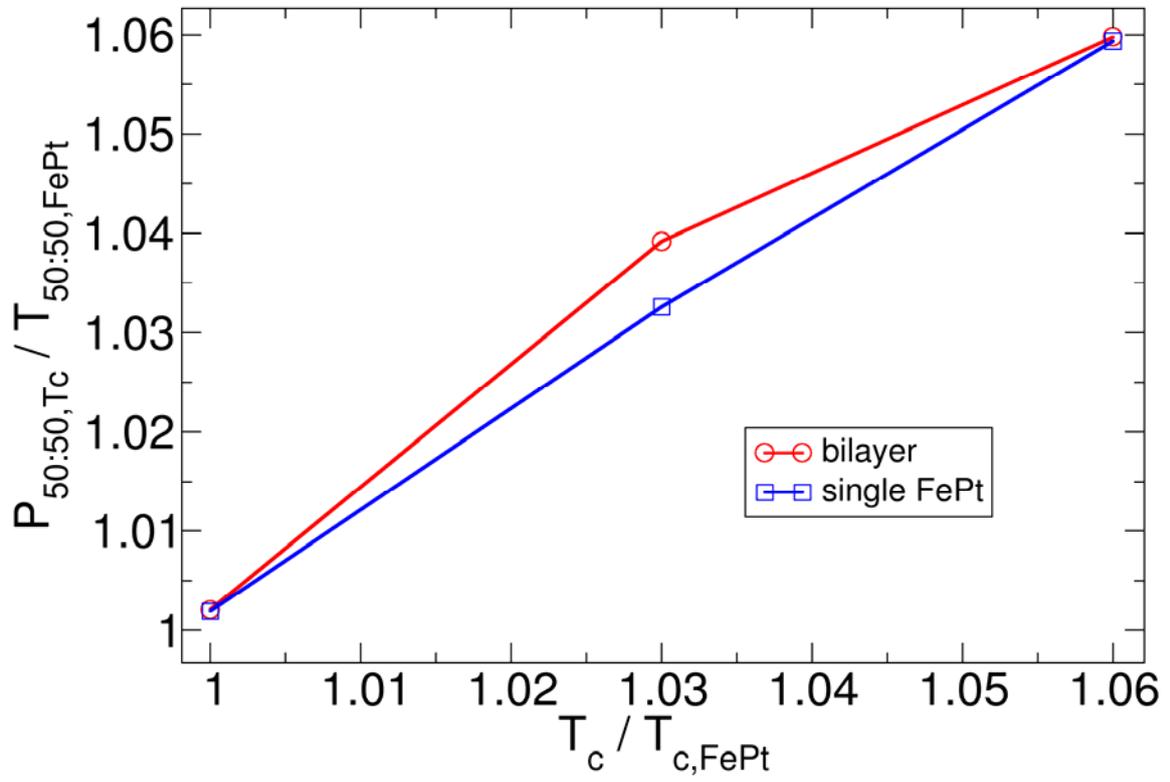

Fig. 12: Comparison of normalized switching temperature $T_{50:50}$ as function of Curie temperature $T_c$.

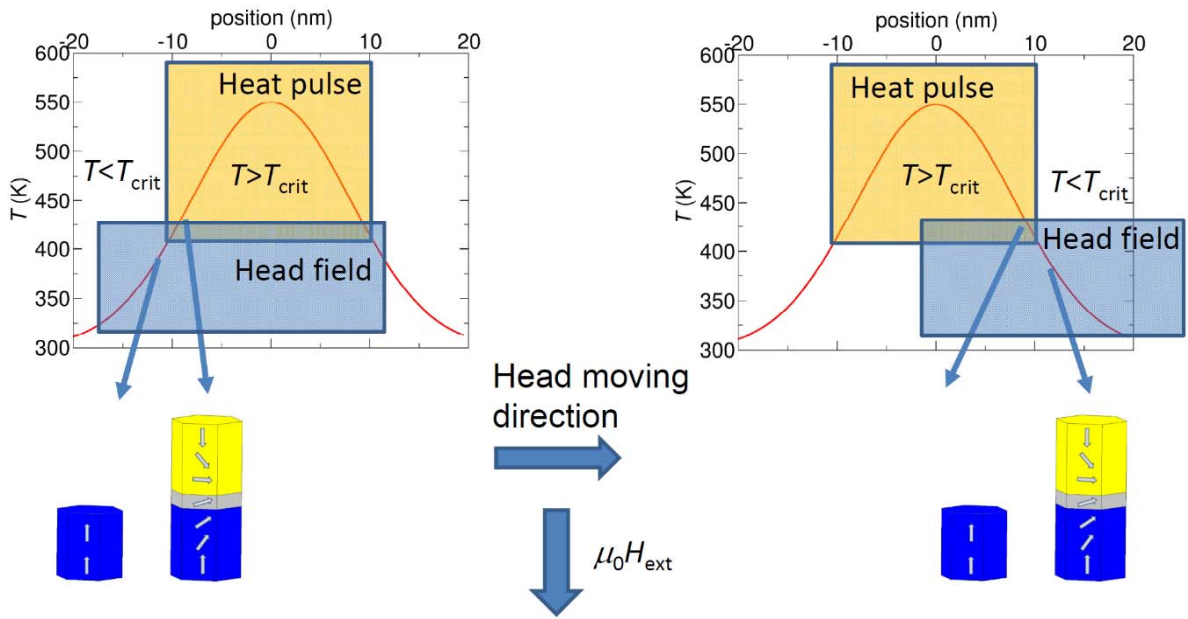

Fig. 13: Concepts of heat assisted / heat protected recording utilizing *first* order phase transition materials.

**Table 1:** Summary of the of the key output parameter: The temperature where 50% of the elements have switched ($T_{50:50}$), the maximum switching probability $P_{max}$, the maximum

slope of $P(t)$, which is $(dP(T)/dT)_{max}$ and the standard deviation of $dP(T)/dT$ which is $\sigma_{dP/dT}$ are shown.

|  | FePt 1 | FePt 2 | FePt 3 | Bilayer 1 | Bilayer 2 | Bilayer 3 | Bilayer 4 | Bilayer 5 | Bilayer 6 |
|---|---|---|---|---|---|---|---|---|---|
| $d$ [nm] | 5 | 10 | 5 | 5 | 5 | 5 | 5 | 5 | 10 |
| $h$ [nm] | 10 | 10 | 30 | 10 | 10 | 10 | 10 | 10 | 10 |
| $J_{s,soft}$ [T] | - | - | - | 1.4 | 2.1 | 1.4 | 1.4 | 2.1 | 1.4 |
| $A_{int}$ | - | - | - | full | full | 1/4 | full | full | Full |
| $T_{c,soft}/T_{c,hard}$ | - | - | - | 1.4 | 1.4 | 1.4 | 1.4 | 1 | 1.4 |
| $t_{hard}/t_{soft}$ | - | - | - | 1 | 1 | 1 | 3 | 1 | 1 |
| $T_{50:50}$ [K] | 524 | 541 | 529 | 488 | 460 | 481 | 515 | 473 | 522 |
| $P_{max}(T)$ | 0.88 | 0.98 | 0.97 | >1-10$^{-4}$ | >1-10$^{-4}$ | 0.99 | 0.98 | 0.97 | >1-10$^{-4}$ |
| $(dP(T)/dT)_{max}$ [1/T] | 0.040 | 0.1 | 0.046 | 0.016 | 0.017 | 0.017 | 0.022 | 0.017 | 0.020 |
| $\sigma_{dP/dT}$ [K] | 8.7 | 3.8 | 8.34 | 23.4 | 21.9 | 23.4 | 17.4 | 22.8 | 19.0 |

**Table 2:** Summary of the estimated jitter values for single FePt elements and bilayer element with $d$=5nm and a height of $h$ = 10 nm.

|  | FePt 1 | Bilayer 1 |
|---|---|---|
| $\sigma_{x,dP/dT}$ (nm) | 0.5 | 1.19 |
| $\sigma_{x,d}$ (nm) | 0.6 | 0.93 |